# S1-MatAgent: A planner driven multi-agent system for material discovery

Xinrui Wang[1, 2]†, Chengbo Li[3, 4]†, Boxuan Zhang[1, 2], Jiahui Shi[1, 2]*, Nian Ran[3, 4]*, Linjing Li[1, 2], Jianjun Liu[3, 4, 5], Dajun Zeng[1, 2]

**Affiliations:**

[1]State Key Laboratory of Multimodal Artificial Intelligence Systems, Institute of Automation, Chinese Academy of Sciences, Beijing 100190, China.

[2]School of Artificial Intelligence, University of Chinese Academy of Science, Beijing 100049, China.

[3]State Key Laboratory of High Performance Ceramics, Shanghai Institute of Ceramics, Chinese Academy of Sciences, Shanghai 200050, China.

[4]Center of Materials Science and Optoelectronics Engineering, University of Chinese Academy of Sciences, Beijing 100049, China.

[5]School of Chemistry and Materials Science, Hangzhou Institute for Advanced Study, University of Chinese Academy of Sciences, Hangzhou 310024, China.

†These authors contributed equally to this work.

*Corresponding author. Email: N.R. (rannian@mail.sic.ac.cn), J. S. (jiahui.shi@ia.ac.cn)

**Abstract:** The discovery of high-performance materials is crucial for technological advancement. Inverse design using multi-agent systems (MAS) shows great potential for new material discovery. However, current MAS for materials research rely on predefined configurations and tools, limiting their adaptability and scalability. To address these limitations, we developed a planner driven multi-agent system (S1-MatAgent) which adopts a Planner-Executor architecture. Planner automatically decomposes complex materials design tasks, dynamically configures various tools to generate dedicated Executor agents for each subtask, significantly reducing reliance on manual workflow construction and specialized configuration. Applied to high-entropy alloy catalysts for hydrogen evolution reactions in alkaline conditions, S1-MatAgent completed full-cycle closed-loop design from literature analysis and composition recommendation to performance optimization and experimental validation. To tackle the deviations between designed materials and target, as well as high experimental verification costs, S1-MatAgent employs a novel composition optimization algorithm based on gradients of machine learning interatomic potential, achieving 27.7 % improvement in material performance. S1-MatAgent designed 13 high-performance catalysts from 20 million candidates, with $Ni_4Co_4Cu_1Mo_3Ru_4$ exhibiting an overpotential of 18.6 mV at 10 mA cm$^{-2}$ and maintaining 97.5 % activity after 500 hours at 500 mA cm$^{-2}$. The universal MAS framework offers a universal and scalable solution for material discovery, significantly improving design efficiency and adaptability.

**Introduction**

The discovery of high-performance and functional new materials plays a critical role in driving technological progress and industrial development[1-5]. However, traditional materials research and development face challenges such as long experimental cycles, lack of theoretical guidance in research direction, and methodological uncertainties[6,7]. High-throughput computational frameworks[8] accelerate material discovery by screening promising candidate material spaces using open materials databases[9,10], density functional theory[11], and machine learning[12,13]. However, their reliance on expensive computing resources severely restricts the research and development efficiency. Under this background, inverse design[14,15], with the concept of "demand-driven", leverages technologies such as generative models[16,17] and large language models (LLMs) providing a novel approach by directly generating materials that meet specific property requirements[18,19].

Knowledge-driven LLMs, with their powerful ability to integrate text knowledge offer a novel pathway for materials inverse design and demonstrate unique advantages[20,21]. However, materials design is a complex system engineering that involves multiple levels of decision-making, including demand analysis, composition selection, performance prediction, and verification iteration. These tasks are highly coupled and mutually constraining[22]. Although LLMs possess broad general knowledge, it is difficult for LLMs to independently handle such complex design demands characterized by multiple stages, types, and objectives. This can easily affect the feasibility and reliability of the designed materials. Furthermore, since LLMs cannot directly interact with the physical world. If all candidate materials were to be synthesized and verified one by one by researchers, it would result in extremely high time and economic costs. This creates a significant gap between design stage and verification-optimization stage, limiting the iterative efficiency and overall progress of materials design and optimization. Therefore, how to globally coordinate and dynamically plan multi-stage task sequences, while efficiently evaluating and optimizing the performance of generated materials, has become a key challenge in advancing the application of LLMs in materials inverse design.

With the development of artificial intelligence, LLM-driven multi-agent systems (MAS), leveraging their knowledge comprehension and reasoning capabilities, can dynamically plan tasks based on understanding the complex requirements of materials design[23,24]. These agents flexibly employ various tools, such as literature retrieval, computational simulation, and experimental data processing, to achieve material design and optimization, offering an effective approach for the inverse design of new materials. For example, Ghafarollahi A et al. used computation agents to solve problems requiring detailed atomic simulations in materials design, significantly improving the efficiency of developing and analyzing crystalline materials at the atomic scale[25]. Similarly, Hao Li's team introduced agents to construct machine learning models from experimental data, efficiently revealing ion migration mechanisms in solid-state electrolytes[26]. In addition, various specialized agents, including structure generation agents for new material[27], data extraction and reasoning agents for material data[28], and experimental agents for material synthesis[29], have enabled comprehensive analysis and design of materials by leveraging professional knowledge and versatile tools.

However, these agents currently rely on manually constructed fixed workflows. Due to the coupled multi-stage tasks and complex non-standardized requirements involved in materials inverse design, this manually defined method places high demands on researchers' understanding of the problem. In order to reduce the dependency of workflow creation on the experience of researchers and make the task-solving process more flexible, some efforts have begun to explore approaches for task planning. Adib Bazgir et al. utilizes chain-of-thought technology in material multi-agent systems to generate workflows and accordingly dispatches various predefined functional agents[30]. Lianhao Zhou et al. proposed framework can first decompose tasks into subtask sequences using a large reasoning model and human intuition, and then attempt to generate instrumental code for each subtask[31]. However, these works do not prioritize the hierarchical relationships within workflows during the task planning process, and still rely on predefined or partially predefined agents to execute every subtask, limiting the versatility and applicability of MAS in the field of materials design.

To address the above challenges, we developed a novel planner driven multi-agent system (S1-MatAgent) for material discovery. S1-MatAgent is capable of automatically decomposing complex materials design tasks and dynamically planning their execution. It breaks down the materials design process into multiple collaboratively subtask units, such as literature mining, composition generation, structure construction, property calculation, and optimization adjustment. Based on the stage results, it autonomously adjusts design strategies and execution path. Through multi-agent collaboration mechanism, S1-MatAgent achieves closed-loop design workflow, effectively reducing the difficulties arising from multi-factor coupling. To tackle issues of deviation between design materials and targets caused by the hallucination of LLMs and reasoning errors based on incomplete knowledge, as well as high cost of experimental verification, S1-MatAgent integrates lightweight computation agents and materials optimization agents. By leveraging tools such as machine learning interatomic potential (MLIP) and performance descriptors, it enables rapid performance evaluation and iterative optimization of candidate materials. This approach effectively compensates for the limitations of LLMs in physical perception and verification optimization, thereby enhancing the reliability and iterative efficiency of materials design.

We validated the performance of S1-MatAgent by applying it to the inverse design of high-entropy alloy (HEA) catalyst for alkaline hydrogen evolution reaction (HER), a challenging task characterized by data scarcity and a vast space of element combinations. S1-MatAgent successfully designed 13 high-performance catalysts from a design space comprising 20 million candidate catalysts. Experimental results demonstrated that S1-MatAgent designed novel catalyst $Ni_4Co_4Cu_1Mo_3Ru_4$ exhibits an overpotential of 18.6 mV at a current density of 10 mA cm$^{-2}$ and maintained 97.5 % of its performance after 500 hours HER at 500 mA cm$^{-2}$. S1-MatAgent not only effectively addresses the challenges of multi-task coupling and verification optimization in materials inverse design, but also demonstrates high efficiency and reliability in practical materials discovery. It provides a universal solution for materials inverse design.

**Results**

**Architecture of S1-MatAgent and related tool chains**

Leveraging the extensive world knowledge and advanced tool-calling capabilities of LLMs, agents built upon these models have exhibited a degree of versatility and efficiency in materials science that surpasses human experts. In current MAS designed for materials research, each agent's profile, functions, and tools are typically rigidly predefined. These functionally specialized agents operate as fixed "processes" within a manually configured workflow. While effective for narrow tasks, this architecture, particularly in complex applications such as materials reverse design, heavily depends on the researchers' preconceived understanding of task decomposition. As a result, it considerably restricts the MAS's capacity for exploratory problem-solving and limits its adaptability and scalability to new tasks.

Unlike the traditional agent collaboration framework where researchers manually define task links and configure interaction logic among multiple agents, S1-MatAgent adopts a Planner-Executor architecture which incorporates dynamic workflow generation and adaptive agent configuration. Acting as a central scheduler, the Planner interacts directly with the user to receive a root task, automatically builds task workflows, and automatically generates several Executors to solve each subtask. Executor here refers to a task-specific customized agent, which incorporates unique system messages and a dedicated toolset.

The Planner of S1-MatAgent uses an LLM-based planning agent, which manages the autonomous, dynamic development of task workflows. Upon receiving a user request, Planner first analyzes the available tools to determine whether the task can be solved directly. For complex problems that require decomposition, Planner breaks them down into a set of parallel or sequential subtasks by constructing a hierarchical task network (HTN). For each compound task in HTN, it will be decomposed into subtasks until the task is primitive. This decomposition results in a tree structure that captures both task hierarchy and temporal dependencies, which is systematically recorded in a working memory file. Based on the dependency relationships within the tree, Planner constructs an executable workflow and assigns dedicated agents as Executors, equipped with tailored profile and toolset, to address each active subtask. As Executors return their results, Planner updates working memory and autonomously triggers subsequent subtasks. Once all subtasks under a parent task are completed, Planner aggregates their outcomes into a comprehensive final result (Fig. 1a).

Figure 1b shows the complete HTN by breaking down the material design task: "Design highly active high-entropy alloy catalysts for hydrogen evolution reaction in an alkaline environment based on literatures". The toolset provided includes tools for literature analysis, coding, HEA recommendation, and HEA optimization, et al. The root task is structured into two phases and four subtasks: first, extraction and statistical analysis of components based on literature information (Subtask 1, Fig. 1b), followed by component recommendation and optimization (Subtask 2, Fig. 1b).

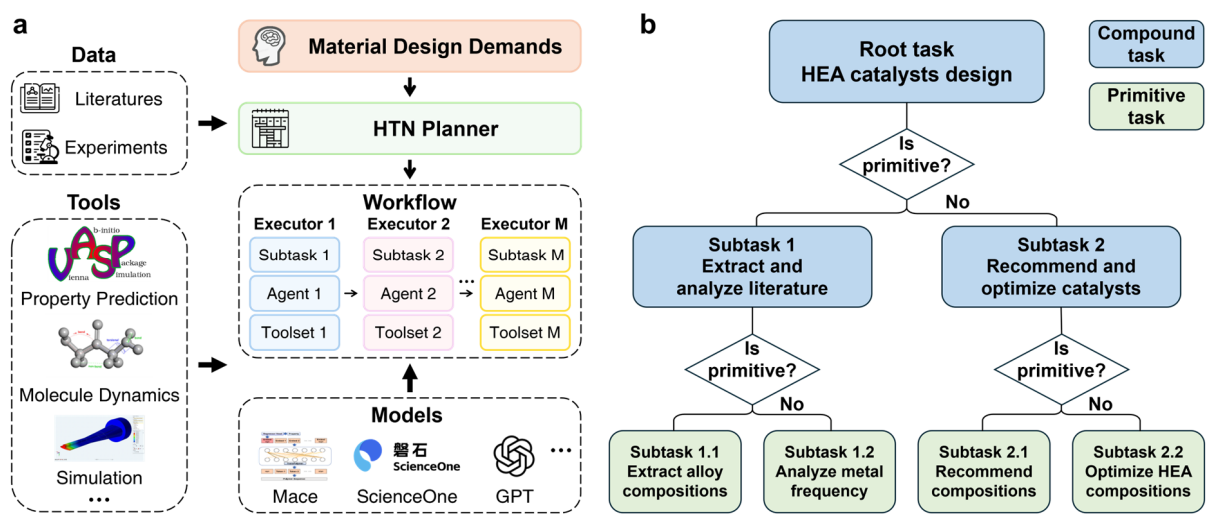

**Fig.1 | S1-MatAgent: A planner driven multi-agent dynamic collaborative system for material reverse design. a** The planner's dynamic creation of a hierarchical task decomposition tree, building of workflows, and dynamic configuration of agents to perform primitive tasks. **b** Task decomposition tree diagram of catalysts design task.

Figure 2 shows all Executors of subtasks in the HEA design task. Executor 1.1 uses Literature Extraction Tool to extract potential chemical formulas from text using methods such as element name unification and regular expression matching. Executor 1.2 uses Metal Element Frequency Statistics Tool to automatically write and execute code using a LLM to count the frequencies of metal elements in a batch of chemical formulas and plot a frequency histogram. Executor 2.1 uses HEA Recommendation Tool to recommend potential high-activity HEA along with justification and references. Recommendation is completely based on a comprehensive analysis on given literatures by the ScienceOne model[32]. After that, all recommended compositions are integrated, illegal compositions are filtered out, and formatted recommendations are generated. Executor 2.2 uses HEA Optimization Tool, based on mechanism modeling and MLIP gradients, iteratively optimizes previously recommended components until activity no longer improves.

During the literature information processing phase, S1-MatAgent utilized a dataset comprising 1,231 relevant articles from the past five years. This dataset, along with the root task, was provided to the MAS. Following the pre-generated workflow, the agents of Executors extract catalyst chemical formulas, and calculate the frequencies of metal elements (Fig. 2a-d). Based on the extraction and statistical analysis, among the 25 metal elements mentioned in the literature, the ten most frequent metal elements identified are Ni, Pt, Co, Fe, Mo, Ru, Cu, Ir, Pd, and Rh (Fig. 2c and d). By extracting chemical formulas and counting the frequencies of the metal elements within them, subsequent component recommendation and optimization steps are focused on elements with high potential for forming highly active catalysts. This approach significantly narrows down the search space for HEA compositions, reducing the vast design space of 20 million possible five-element combinations from 25 metal elements into a more manageable set of nickel-based compositions combined with four other high-probability metal elements. In the composition recommendation phase, the subtask Executor employs the ScienceOne model to perform a comprehensive comparison of information extracted from the literature source. It generates multiple candidate recommendations and automatically writes code to screen the legitimacy of the proposed components. These filtered several recommendations serve as potential compositions for highly active high-entropy alloy catalysts under alkaline conditions, and will undergo further activity optimization in the subsequent workflow (Fig. 2e and f). Then, a verification and optimization process of recommended compositions is conducted, forming the final candidate HEA for selection and synthesis by experimenters (Fig. 2g and h).

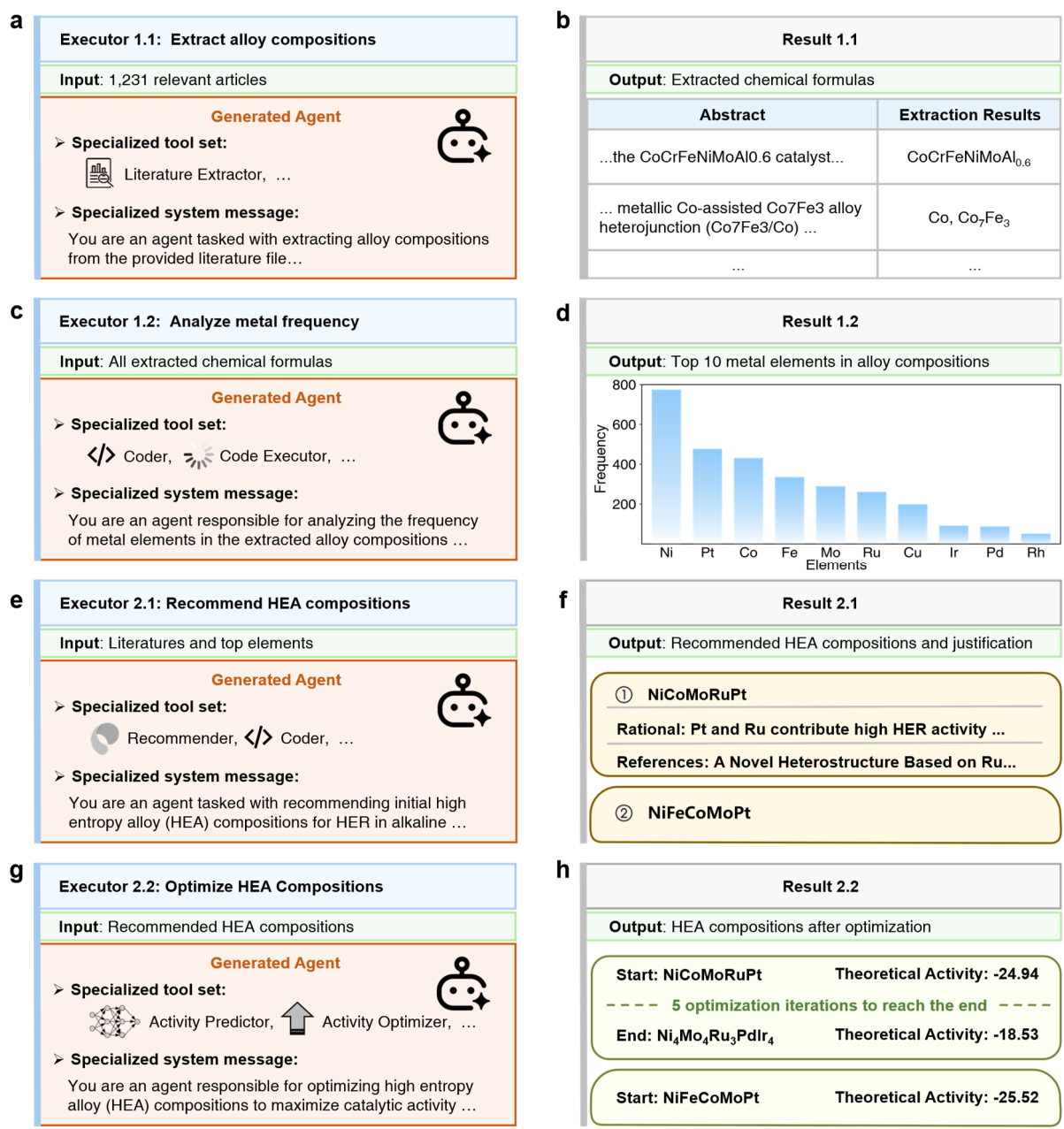

**Fig.2 | Execution process of all subtasks in the HEA design task.** The Executor configuration and execution results for the task of **a-b** alloy composition extraction, **c-d** metal frequency analyzing, **e-f** HEA composition recommendation, and **g-h** HEA compositions optimization for better HER activity.

**Evaluation and optimization of HEA compositions**

In order to evaluate and optimize the activity of these candidate materials, the agent in the material optimization task invokes a catalyst performance evaluation and component optimization tools to iteratively refine each potential composition. This process yields a set of recommended component configurations with improved theoretical activity, which are then provided to experimenters for experimental synthesis and performance evaluation.

To enable S1-MatAgent to rapidly evaluate the performance of designed catalysts from theoretical perspective, we provided a HEA structure generation tool and established a HER activity evaluation model based on the Volmer-Tafel mechanism in combination with MLIP (Fig. 3a). Specifically, we constructed a dataset containing 43,064 HEA structures, based on which a MACE model[33] was fine-tuned to achieve precise simulation of the HER process of HEA. After fine-tuning, the RMSE for energy prediction of this model on the test set is only 3.3 meV per atom (Fig. 3b). The rapid quantification of the HER activity of the catalyst was achieved by constructing HEA structure containing reaction intermediates using structure generation tool and calculating the water adsorption energy, water dissociation energy barrier and hydrogen adsorption-free energy during the HER process by the MACE model.

Due to hallucination of LLMs, reasoning biases based on incomplete or incorrect knowledge, the catalysts designed may have deviations between their actual performance and expected results. To address this, we provide a catalyst optimization algorithm for S1-MatAgent to dynamically adjust the catalyst and optimize its performance. This optimization algorithm achieves synchronous optimization of material element types and composition ratios based on the gradient of the fine-tuned MACE model. Based on the computational graph of the MACE model, we can compute the gradient of the model's HER activity output with respect to the elemental representation of the atomic configuration. This allows us to derive the gradient of the HER activity with respect to the same elemental representation. Through this approach, the contribution of each element to the HER activity, whether enhancing or diminishing, can be quantitatively assessed. We propose that for HEA, enhancing the positive contribution of each element to the HER activity will facilitate the design of highly active catalysts for HER. After the model predicts the HER activity of the catalyst, the optimization algorithm adjusts both the element composition ratios and replaces the element types sequentially according to the gradient magnitude of each element with respect to the HER activity. After each optimization step, the algorithm comprehensively evaluates all candidate materials and selects new materials with superior performance for the next iteration. This process continues until the HER activity of the material no longer significant increases (Fig. 3c). We used this optimization algorithm to optimize the HER activity of 400 HEA catalysts. After optimization, the average activity improved from -21.12 to -15.28, representing a significant increase of 27.7 % (Fig. 3d).

To compare the effectiveness of the gradient-based algorithm with the traditional optimization algorithm, such as genetic algorithm[34], we constructed a population of eight components and applied the genetic algorithm to the population, compared to applying the gradient-based optimization algorithm to each individual in the population. After 10 iterations, our algorithm increased the highest HER activity of a population with an average final activity 2.4 times higher than that of the genetic algorithm and increase in the first round being 2.8 times greater. Considering the number of component evaluations, we equate one iteration of gradient-based optimization for a single component to one iteration of genetic algorithm. The median activity evaluation value of individuals in the population is plotted against the number of iterations during gradient-based optimization. The results show that, even when starting from a single component, more than half of the components achieve higher activity than those optimized via the genetic algorithm, demonstrating the significant superiority of the gradient-based algorithm (Fig. 3d).

Within a design space of 20 million candidate materials composed of 25 metal elements for five-element HEA, ScienceOne model designed 13 highly active candidate materials. After iterative optimization of these candidate materials through optimization algorithm, we selected five materials $Ni_4Mo_4Ru_4Pd_1Pt_3$, $Ni_5Fe_3Mo_4Ru_1Pd_3$, $Ni_4Mo_4Ru_3Rh_1Pt_4$, $Ni_4Fe_4Mo_4Pd_1Pt_3$, and $Ni_4Co_4Cu_1Mo_3Ru_4$ suitable for laboratory synthesis for experimental verification.

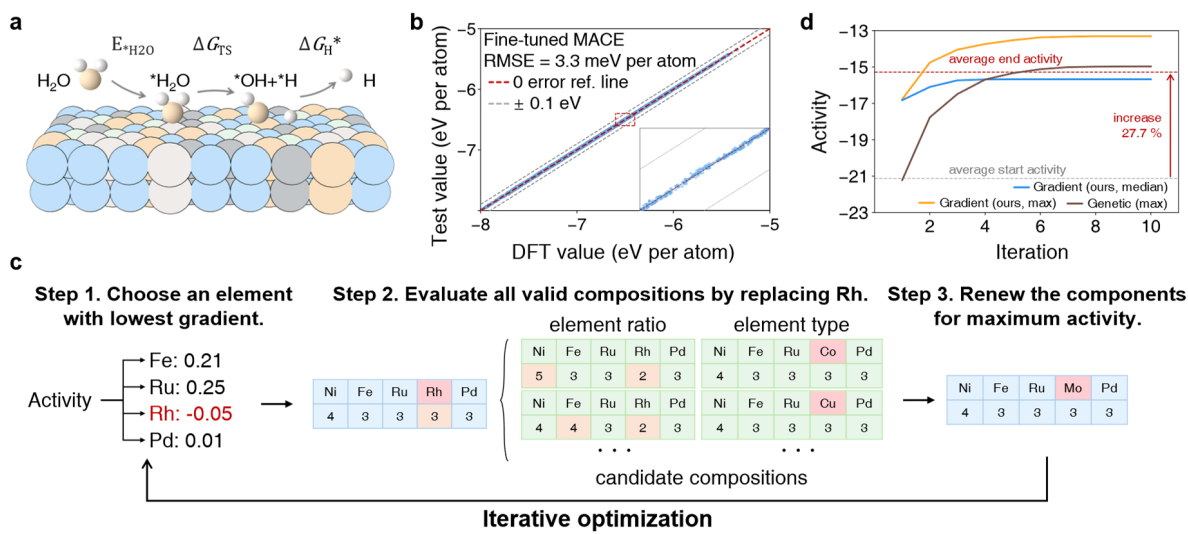

**Fig.3 | Evaluation and Optimization of HER Activity in HEA. a** Schematic diagram of HER activity assessment under alkaline conditions. **b** Performance of the fine-tuned MACE model for energy prediction on the test set with partial enlarged region inserted. **c** One-step optimization process of the gradient-based optimization algorithm. **d** Performance comparison of different component optimization algorithms. The solid lines represent the activity optimization process of the gradient-based algorithm and genetic algorithm for the same population of eight compositions. The dashed line indicates the average activity before and after applying the gradient-based algorithm to all individual compositions.

**Experimental verification**

Figure 4 shows the electrochemical performance and morphology characterization. The LSV curve (Fig. 4a and b) shows that $Ni_4Co_4Cu_1Mo_3Ru_4$ has the most excellent HER activity, achieving overpotentials of 18.6 mV at 10 mA cm$^{-2}$, 93.5 mV at 100 mA cm$^{-2}$, and 146.4 mV at 500 mA cm$^{-2}$. In addition, $Ni_4Co_4Cu_1Mo_3Ru_4$ also has a low Tafel slope (65.6 mV dec$^{-1}$), reflecting its outstanding HER kinetics (Fig. 4c). Nyquist plots show that $Ni_4Co_4Cu_1Mo_3Ru_4$ has the lowest charge transfer resistance among its counterparts, which is only 2.51Ω (Fig. 4d), further indicating the faster kinetics process for producing $H_2$. This provides the foundation for outstanding dynamics. The $C_{dl}$ values of $Ni_4Co_4Cu_1Mo_3Ru_4$ is 121.9 mF cm$^{-2}$, reflecting a high electrochemical active surface area (Fig. 4e). Moreover, it exhibits excellent stability, retaining 97.5 % of its activity after 500h at 500 mA cm$^{-2}$ (Fig. 4f). $Ni_4Co_4Cu_1Mo_3Ru_4$ presents a particle morphology (Fig. 4g). The SEM image of the low-magnification overview and the corresponding elemental mapping results indicate the presence and uniform distribution of Ni, Co, Cu, Mo, Ru elements, confirming that the elemental composition of the synthesized material is consistent with that of the designed material (Fig.4h). The above characterization and electrochemical test results confirm the precise synthesis of the designed sample and its excellent HER activity, verifying the feasibility of S1-MatAgent.

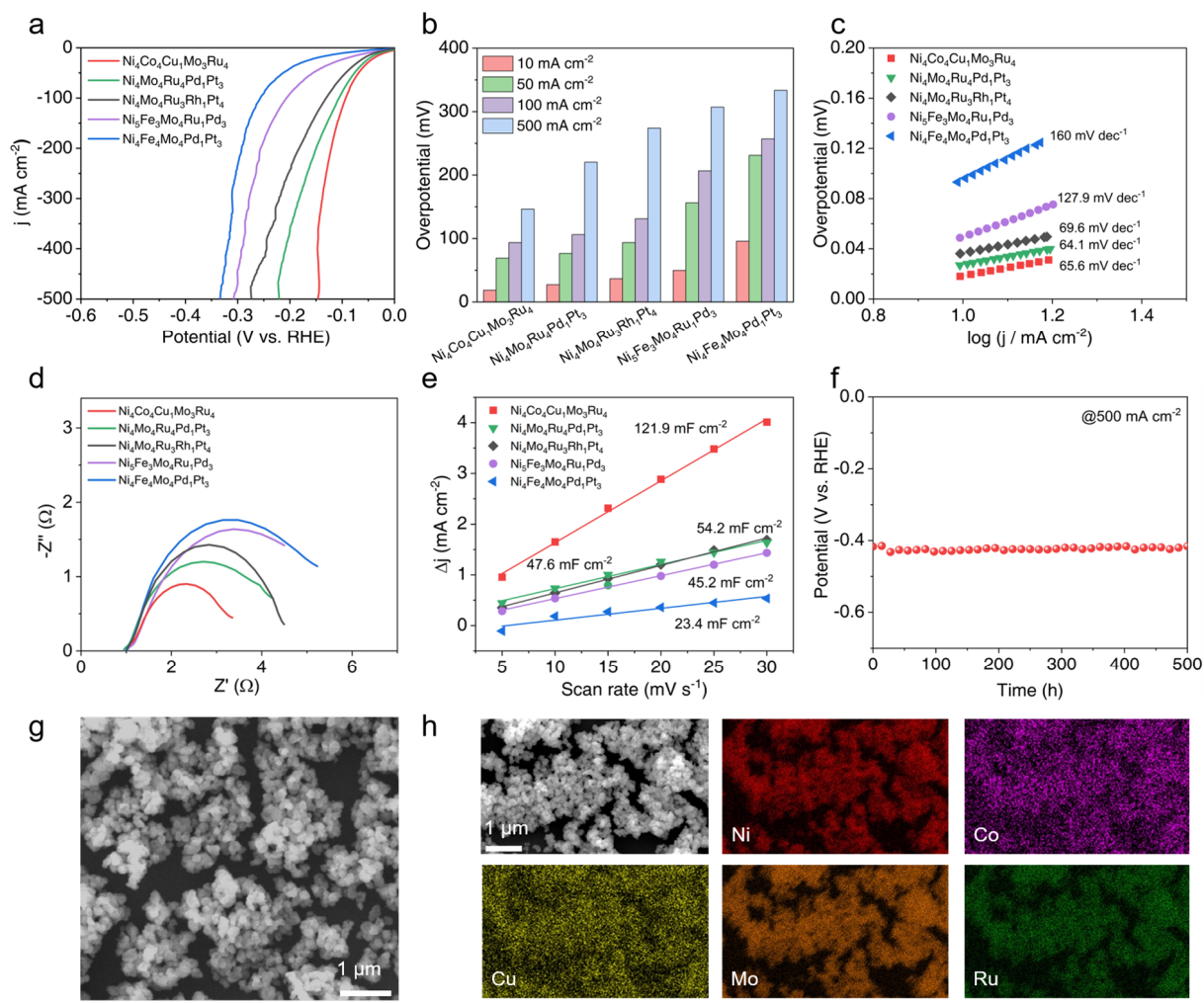

**Fig.4 | Characterizations and alkaline HER performance. a,** LSV curves, **b,** overpotential values, **c,** Tafel values, **d-e,** EIS curves and $C_{dl}$ values of $Ni_4Mo_4Ru_4Pd_1Pt_3$, $Ni_5Fe_3Mo_4Ru_1Pd_3$, $Ni_4Mo_4Ru_3Rh_1Pt_4$, $Ni_4Fe_4Mo_4Pd_1Pt_3$, and $Ni_4Co_4Cu_1Mo_3Ru_4$. **f,** Stability test of $Ni_4Co_4Cu_1Mo_3Ru_4$ at 500 mA cm$^{-2}$ for 500 hrs.

**g-h,** SEM image and elemental mapping of $Ni_4Co_4Cu_1Mo_3Ru_4$.

**Discussion**

This study proposes a planner driven multi-agent system (S1-MatAgent) to address the core challenges faced by LLMs in materials inverse design, such as multi-stage task coupling, lack of domain knowledge, and limited verification and optimization capabilities. Based on a Planner-Executor architecture, S1-MatAgent enables automatic decomposition and dynamic planning of complex material design tasks, significantly reducing reliance on manual workflow construction and specialized configuration, while enhancing versatility and adaptability in materials design. S1-MatAgent is highly scalable, and more tools can be added as needed and more functions can be freely configured. S1-MatAgent incorporates a gradient-based optimization algorithm designed to improve material performance. This algorithm can achieve a 27.7 % enhancement of designed materials, effectively compensating for the deficiencies of LLMs in material verification and closed-loop optimization. In the case on designing HEA catalysts for alkaline HER, S1-MatAgent successfully achieved full-cycle closed-loop design from literature analysis and component recommendation to performance optimization and experimental verification. Among the 13 high-performance materials designed from 20 million design spaces, the designed $Ni_4Co_4Cu_1Mo_3Ru_4$ catalyst exhibited an overpotential of 18.6 mV at a current density of 10 mA cm$^{-2}$ and maintained 97.5 % of its HER performance after 500 hours HER at 500 mA cm$^{-2}$.

This research provides a universal intelligent agent framework for materials inverse design of complex material systems, promoting paradigm shift in AI-driven materials research from single-point assistance through manual configuration to full-process autonomous planning. Future work will focus on extending S1-MatAgent to diverse multi-component material systems and exploring seamless integration with automated experimental platforms to enable higher-throughput material design and discovery.

## Methods

**HER mechanism modeling and activity descriptors**

To evaluate the catalytic activity of HEA components in the HER under alkaline conditions during the component optimization phase, we constructed a HER activity descriptor. Based on the Volmer-Tafel mechanism, we modeled the HER of the HEA as a three-step reaction: water adsorption, water dissociation, and the Tafel reaction. By calculating the enthalpy changes or kinetic barriers of these three elementary reactions, we can assess the ease with which these three steps occur. In alkaline environments, the protons for HER originate from water dissociation, which adds an additional energy barrier and significantly reduces the reaction kinetics. Therefore, during the component optimization phase, we used the inverse of the water dissociation kinetic barrier as the activity descriptor and optimized candidate highly active components by maximizing this activity descriptor. After the optimization algorithm converged, the multi-agent system provided feedback on the final list of potential highly active components, which were then screened by experts using a combination of water adsorption enthalpy changes, Tafel reaction enthalpy changes, and expert experience.

For water adsorption on slabs, the enthalpy change calculations for the Tafel reaction, and the kinetic barrier calculations for water dissociation, we perform calculations for all possible adsorption sites on the surface and take the average of the calculated values for all sites as the calculation result for that substrate. The kinetic barrier is approximated using the structural difference method, which is the energy difference between the highest energy structure and the initial structure.

**Construction of HEA MLIP model**

In recent years, methods based on machine learning algorithms to construct potential models, such as MACE, have demonstrated remarkable superiority in predicting interatomic interaction potentials due to their high efficiency and accuracy. We constructed a dataset of structures of HEA slabs and their adsorbed intermediates to fine-tune the MACE model. Specifically, the dataset covers slab structures and their adsorbed intermediates, constructed with for every elemental quinary combination across a range of allowed compositional ratios. Ultimately, 630 trajectories and 43,064 configurations with energy and force labels were obtained.

We used this dataset to perform multi-head fine-tuning on MACE to prevent catastrophic forgetting. The fine-tuned potential model achieved an RMSE of 3.3 meV per atom on the energy of the test set.

**Gradient-based Composition Optimization Algorithm**

Previous applications of MLIPs have often been restricted to extracting specific quantitative values, such as energy or force, from the model outputs. These values are then used in downstream tasks, such as high-throughput screening. However, MLIPs inherently construct an end-to-end computational graph that maps atomic configurations to property predictions. Utilizing only the final numerical outputs obtained through forward propagation represents an underutilization of this comprehensive computational structure.

We developed a composition optimization algorithm based on the gradient of a MLIP to perform an efficient heuristic search in the compositions' space. We define a composition as the union of element types and element ratios. At each iteration, we use the MLIP model to obtain the average gradient of the activity descriptor for each element corresponding to the current composition. The element with the lowest gradient is then replaced. Two replacement strategies are considered: changing element ratios, where the content of the element is reduced by one unit and the content of another element is increased by one unit; and changing element types, where all atoms of the element are replaced with atoms of another candidate element. During the entire substitution process, we only consider compositions that preserve the maximum nickel content within the quinary alloy while maintaining elemental ratios between 5 % and 35 %. The activity of all candidate compositions obtained by the two replacement strategies is evaluated sequentially. If the optimal activity among the candidate compositions is better than that of the current composition, the current composition is replaced by this candidate. The composition optimization algorithm starts from an initial composition and proceeds through successive iterations of optimization until no candidate composition outperforms the current one, thereby yielding the final optimized composition.

To demonstrate the superiority of our algorithm, we compared the composition optimization algorithm based on machine learning potential gradients with a genetic algorithm. In each generation of the genetic algorithm, a new population is constructed: the composition with the highest activity from the previous generation is retained, while the rest are generated through mating, crossover, and mutation processes based on parent compositions from the prior generation. Considering that the gradient-based method evaluates approximately 7 candidate compositions per iteration, we set the population size to 8 so that both algorithms assess a similar number of compositions per round.

During the generation of each offspring in a round, parent compositions are selected according to the activity performance of individuals from the previous generation. To ensure that each offspring composition consists of exactly five elements, we first perform element matching, followed by a single-point crossover triggered with an 80 % probability, meaning 1 to 4 elements are exchanged between the parent compositions simultaneously, while ensuring the same two elements are not crossed into the same offspring. The newly exchanged elements are proportionally reintegrated into the remaining elemental ratios of the original composition. A mutation operation is then triggered with a 20 % probability to finally yield one offspring composition.

Mutation occurs with equal likelihood between ratio mutation and element mutation. In ratio mutation, the composition is randomly altered to any valid composition within a Hamming distance of 2 from the original. In element mutation, one non-Ni element is randomly replaced by another candidate element. After constructing the population each round, the genetic algorithm evaluates the activity of each composition as the individual's fitness. The iteration stops when the best fitness in the current generation equals that of the previous generation.

**Materials and reagents**

KOH, Ni(NO$_3$)$_3$·6H$_2$O, Fe(NO$_3$)$_3$·9H$_2$O, Co(NO$_3$)$_2$·6H$_2$O, (NH$_4$)$_2$MoO$_4$·4H$_2$O, K$_3$RuCl$_6$, K$_3$RhCl$_6$, H$_2$PtCl$_6$·6H$_2$O, Na$_2$PdCl$_4$, Cu(NO$_3$)$_2$·3H$_2$O and NaBH$_4$ were brought from Aladdin Chemical Reagent Company (Shanghai, China). All aqueous solutions were prepared using ultrapure water (resistivity = 18.2 MΩ cm, Milli-Q water) throughout the whole experiments.

**Material synthesis**

Preparation of Ni$_4$Mo$_4$Ru$_4$Pd$_1$Pt$_3$, Ni$_5$Fe$_3$Mo$_4$Ru$_1$Pd$_3$, Ni$_4$Mo$_4$Ru$_3$Rh$_1$Pt$_4$, Ni$_4$Fe$_4$Mo$_4$Pd$_1$Pt$_3$, and Ni$_4$Co$_4$Cu$_1$Mo$_3$Ru$_4$: An aqueous solution containing Ni(NO$_3$)$_3$·6H$_2$O, Fe(NO$_3$)$_3$·9H$_2$O, Co(NO$_3$)$_2$·6H$_2$O, (NH$_4$)$_2$MoO$_4$·4H$_2$O, K$_3$RuCl$_6$, K$_3$RhCl$_6$, H$_2$PtCl$_6$·6H$_2$O, Na$_2$PdCl$_4$, and Cu(NO$_3$)$_2$·3H$_2$O with a total metal ion concentration of 320 mM. The concentration of each metal ion is allocated in molar ratio. Additionally, 80 mM NaBH$_4$ was added. After continuous stirring at 500 r min$^{-1}$ for 40 min, the mixture was centrifuged. The resulting sample was subsequently washed three times with ultrapure water and freeze-dried.

**Density functional theory calculations**

All density functional theory (DFT) calculations were performed in the Vienna Ab initio simulation package (VASP)[35]. The generalized gradient approximation (GGA+U)[36] with the Perdew-Burke-Ernzerhof exchange-correlation functional and a 520 eV cutoff for the plane-wave basis set were employed[37]. The corrections of U-J are obtained from the Material Project[38]. The projector-augmented plane wave (PAW) was used to describe the electron-ion interactions[39]. The convergence threshold was set as 10$^{-4}$ eV in energy and 0.01 eV/Å in force, respectively. The empirical dispersions of Grimme (DFT-D3) were applied to account for the long-range van der Waals interactions[40]. The k-point sampling of the Brillouin zone was obtained using a 3 × 3 × 1 grid for the unit by Monkhorst-Pack scheme. A large vacuum slab of 20 Å was inserted in the z direction for surface isolation to prevent interaction between two neighboring surfaces.

## Characterizations

The morphology and structure of these as-synthesized samples were characterized by a field-emission gun scanning electron microscope (SEM, Hitachi S4800) and a transmission electron microscopy (TEM, FEI Tecnai F20). X-ray diffraction (XRD) patterns were recorded by a X-Pert3 powder diffractometer (PANalytical, Holland) with Cu-K$_\alpha$ radiation ($\lambda$=1.5416 Å). X-ray photoelectron spectroscopy (XPS) measurements were performed on an AXIS Supra (Kratos Analytical Ltd. England) with monochromatic 150 W Al-K$_\alpha$ radiation. The molar ratio of elements was measured with ICP-OES (Thermo Fisher iCAP PRO).

## Electrochemical measurements

All electrochemical tests were measured using a three-electrode system on CS310M electrochemical workstation. Disperse 8mg of the sample in 500 μl of ethanol, 470μl of water and 20μl of naphthol. Coat 150μl of the mixed solution on a 1cm×1cm carbon cloth as the working electrode. HER performances were tested by using the pre-prepared working electrode, Hg/HgO (1M KOH) electrode as reference electrode, and graphite rod as counter electrode. The LSV curves were recorded with the scan rate of 2 mV s$^{-1}$. And potentials were corrected with 90% iR-compensation. Electrochemical impedance spectroscopy (EIS) was conducted from 100 kHz to 0.01 Hz. The electrochemical active surface area (ECSA) was estimated from electrochemical double-layer capacitance ($C_{dl}$). And $C_{dl}$ values were determined with typical CV tests at various scan rates (5, 10, 15, 20, 25, 30 mV s$^{-1}$) in the range of non-faradic region.

## Data availability

The data that support the findings of this study are available from the corresponding author upon reasonable request.

## Code availability

The code for S1-MatAgent is provided at https://github.com/ScienceOne-AI/S1-MatAgent.

## Acknowledgments

This work is financially supported by Strategic Priority Research Program of Chinese Academy of Sciences (XDA0480301), Advanced Materials-National Science and Technology Major Project (2025ZD0619502, 2025ZD0613501), National Natural Science Foundation of China, NSFC (22403103), Sponsored by Shanghai Sailing Program(23YF1454900), and the Science and Technology Commission of Shanghai Municipality (25CL2902100).

**Ethics declarations**

The authors declare no competing interests.

**Author contributions**

N.R. and J.S. designed the project. X.W. and C.L. completed the entire project code and experiments. B.Z. performed data analysis. C.L. and X.W. completed the manuscript. J.S., N.R., L.L., J.L. and D.Z revised the manuscript. N.R., J.L. and D.Z supervised the project and provided funding.

**Competing interests**

The authors declare no competing interests.

**Additional information**

**Correspondence and requests for materials** should be addressed to N.R., and J. S.